\documentclass[12pt]{article}
\usepackage{amssymb,amsmath}
\usepackage{natbib}
\usepackage{graphicx}
\usepackage{epsfig}

\begin{document}

\title{On the maximum Lyapunov exponent of the motion in a chaotic layer}

\author{I.\,I. Shevchenko\/\thanks{e-mail:~iis@gao.spb.ru} \\
Pulkovo Observatory RAS, 196140 St.\,Petersburg, Russia}

\date{}

\maketitle

\begin{abstract}
The maximum Lyapunov exponent (referred to the mean half-period of
phase libration) of the motion in the chaotic layer of a nonlinear
resonance subject to symmetric periodic perturbation, in the limit
of infinitely high frequency of the perturbation, has been
numerically estimated by two independent methods. The newly
derived value of this constant is $0.80$, with precision
presumably better than $0.01$.
\end{abstract}


\section{Introduction}
\label{intro}

On the basis of results of extended numerical experiments,
Chirikov~\cite{C78,C79} noted that the maximum Lyapunov exponent,
referred to the mean half-period of phase libration, of the motion
in the chaotic layer of a nonlinear resonance subject to symmetric
periodic perturbation, is approximately constant in a wide range
of a parameter characterizing the perturbation frequency. In this
paper, we estimate the least upper bound for the maximum Lyapunov
exponent of the separatrix map. We show that this bound coincides
with the value of the maximum Lyapunov exponent in the mentioned
problem in the limit of infinitely high frequency of perturbation,
and its value does not depend on the amplitude of the
perturbation, i.~e.\ it is defined robustly. In what follows, this
quantity is called Chirikov's constant. The knowledge of the value
of Chirikov's constant is important for accurate analytical
estimation of Lyapunov exponents in applications in mechanics and
physics~\cite{S00a,S02}.

Nonlinear resonances are ubiquitous in problems of modern
mechanics and physics. Under general
conditions~\cite{C77,C79,LL92}, a model of a nonlinear resonance
is provided by the nonlinear pendulum with periodic perturbations.
The rigid pendulum with the oscillating suspension point is a
paradigm in studies of nonlinear resonances and chaotic behavior
in Hamiltonian dynamics. The Hamiltonian of this system, according
to e.~g.\ \cite{BM95}, is:

\vspace{-3mm}

\begin{equation}
H = {{{\cal G} p^2} \over 2} - {\cal F} \cos \varphi +
    a \left( \cos(\varphi - \tau) + \cos(\varphi + \tau) \right),
\label{h}
\end{equation}

\noindent where $\tau = \Omega t + \tau_0$. The first two terms in
Eq.~(\ref{h}) represent the Hamiltonian $H_0$ of the unperturbed
pendulum, while the two remaining ones the periodic perturbations.
The variable $\varphi$ is the pendulum angle (this angle measures
deviation of the pendulum from the lower position of equilibrium),
and $\tau$ is the phase angle of perturbation. The quantity
$\Omega$ is the perturbation frequency, and $\tau_0$ is the
initial phase of the perturbation; $p$ is the momentum; ${\cal
F}$, ${\cal G}$, $a$ are constants. In what follows, it is assumed
that ${\cal F} > 0$, ${\cal G} > 0$.

Chirikov~\cite{C77,C79} derived the so-called separatrix map
describing the motion in the vicinity of the separatrices of
Hamiltonian~(\ref{h}):

\begin{eqnarray}
& & w_{i+1} = w_i - W \sin \tau_i,  \nonumber \\
& & \tau_{i+1} = \tau_i +
                 \lambda \ln {32 \over \vert w_{i+1} \vert}
                 \ \ \ (\mbox{mod } 2 \pi),
\label{sm}
\end{eqnarray}

\noindent where $w$ denotes the relative (with respect to the
separatrix value) pendulum energy $w = {H_0 \over {\cal F}} - 1$,
and $\tau$ retains its meaning of the phase angle of perturbation.
Constants $\lambda$ and $W$ are parameters: $\lambda$ is the ratio
of $\Omega$, the perturbation frequency, to $\omega_0 = ({\cal F
G})^{1/2}$, the frequency of the small-amplitude pendulum
oscillations; and

\vspace{-3mm}

\begin{equation}
W = {a \over {\cal F}} \lambda (A_2(\lambda) + A_2(-\lambda)) =
 4 \pi {a \over {\cal F}} \lambda^2 \mbox{csch}{\pi \lambda \over 2},
\label{W}
\end{equation}

\noindent
where

\vspace{-3mm}

\begin{equation}
A_2(\lambda) = 4 \pi \lambda {\exp{\pi \lambda \over 2}
\over \sinh (\pi \lambda)}
\label{A2}
\end{equation}

\noindent is the Melnikov--Arnold integral~\cite{C79,LL92,S98a}.
One iteration of map~(\ref{sm}) corresponds to one period of the
pendulum rotation, or a half-period of its libration. The motion
of system~(\ref{h}) is mapped by Eqs.~(\ref{sm})
asynchronously~\cite{S98a}: the action-like variable $w$ is taken
at $\varphi = \pm \pi$, while the perturbation phase $\tau$ is
taken at $\varphi = 0$. The desynchronization can be removed by a
special procedure~\cite{S98a,S00}.

An equivalent form of Eqs.~(\ref{sm}), used e.~g.\ in~\cite{CS84},
is

\begin{eqnarray}
     y_{i+1} &=& y_i + \sin x_i, \nonumber \\
     x_{i+1} &=& x_i - \lambda \ln \vert y_{i+1} \vert + c
                   \ \ \ (\mbox{mod } 2 \pi),
\label{sm1}
\end{eqnarray}

\noindent where $y = {w \over W}$, $x = \tau + \pi$; and the new
parameter

\begin{equation}
c = \lambda \ln {32 \over \vert W \vert}.
\label{c}
\end{equation}

\noindent The standard map represents linearization of the
separatrix map in the action-like variable $y$ near a fixed point;
it is given by the equations

\begin{eqnarray}
     y_{i+1} &=& y_i + K \sin x_i \ \ \ (\mbox{mod } 2 \pi), \nonumber \\
     x_{i+1} &=& x_i + y_{i+1} \ \ \ (\mbox{mod } 2 \pi),
\label{stm}
\end{eqnarray}

\noindent where $K$ is the so-called stochasticity
parameter~\cite{C78,C79}.

\section{The Lyapunov exponents and the dynamical entropy}
\label{lede}

The calculation of the Lyapunov characteristic exponents (LCEs) is
one of the most important tools in the study of the chaotic
motion. The LCEs characterize the rate of divergence of
trajectories close to each other in phase space. A nonzero LCE
indicates chaotic character of motion, while the maximum LCE equal
to zero is the signature of regular (periodic or quasi-periodic)
motion. The quantity reciprocal to the maximum LCE characterizes
the motion predictability time.

Let us consider two trajectories close to each other in phase
space. One of them we shall refer to as {\it guiding} and the
other as {\it shadow}. Let $d(t_0)$ be the length of the
displacement vector directed from the guiding trajectory to the
shadow one at an initial moment~$t = t_0$. The LCE is defined by
the formula~\cite{LL92}:

$$L=\limsup_{{t \to \infty} \atop {d(t_0) \to 0}}
         {1 \over {t-t_0}} \ln{d(t) \over d(t_0)} \, . $$

\noindent In the case of a Hamiltonian system, the quantity $L$
may take $2N$ different values (depending on the direction of the
initial displacement), where $N$ is the number of degrees of
freedom; the LCEs divide into pairs: for each $L_k > 0$ there
exists $L_{k+N} = - L_k < 0$, $k=1,\ldots,N$.

The LCEs are closely related to the dynamical
entropy~\cite{P76,BGS76,C78,C79,M92}. For the Hamiltonian systems
with 3/2 and 2 degrees of freedom, Benettin {\it et al.} proposed
the relation $h = L \mu $~\cite[Eq.~(6)]{BGS76}, where $h$ is the
dynamical entropy, $L$ is the maximum LCE, and $\mu$ is the
relative measure of the connected chaotic domain where the motion
takes place. This formula is approximate. Benettin {\it et
al.}~\cite{BGS76} applied it in a study of the chaotic motion of
the H\'enon--Heiles system.

In what follows, our numerical data is presented on the measure
$\mu$ of the main connected chaotic domain in phase space of the
standard map, the maximum LCE $L$, and the product of $\mu$ and
$L$ for motion in this domain. Two methods for computation of the
chaotic domain measure $\mu$ are used. A traditional ``one
trajectory method'' (OTM) consists in computing the number of
cells explored by a single trajectory on a grid exposed on phase
plane. A ``current LCE segregation method'' (CLSM) is based on an
analysis of the differential distribution of the computed values
of the Lyapunov exponents (current LCEs) of a set of trajectories
with starting values on a grid on phase plane. Both methods were
proposed and used by Chirikov~\cite{C78,C79} in computations of
$\mu$ for the standard map. Analogous methods were used
in~\cite{SM03} in computations of chaotic domain measure in the
H\'enon--Heiles problem.

Fig.1 illustrates discontinuity of the obtained $\mu(K)$ function.
The curve in Fig.1 is obtained by the OTM on the grid $2000 \times
2000$ pixels on phase plane ($x$, $y$) $\in [0, 2 \pi] \times [0,
2 \pi]$, the number of iterations $n_{it} = 10^8$. A prominent
bump in the dependence, shown in detail in Fig.1b, is conditioned
by the process of desintegration of the half-integer resonance in
the course of a sequence of period-doubling bifurcations while $K$
is increasing from $\approx 2$ to $\approx 2.5$. A similar but
less pronounced bump is seen in Fig.1a in the range $4 < K < 4.5$;
this one is due to bifurcations of the integer resonance.

In general, at these moderate values of $K$ (at $K<6$), the
discontinuities are conditioned by the process of absorption of
minor chaotic domains by the main chaotic domain, while $K$
increases.

In Fig.2, we give the plot of the maximum LCE and the dynamical
entropy in a broad range of $K$. Each value of $L$ in Fig.2
represents the mean value of the maximum LCEs over 10 trajectories
of length $n_{it} = 10^7$ each. The initial data for these
trajectories slightly differ from each other, but in all cases are
chosen to lie inside the main chaotic domain. Everywhere in this
work (including the case of the separatrix maps considered below)
the presented values of LCEs have been computed by the tangent map
method described e.~g.\ in~\cite{C78,C79}.

The corresponding dependence of $h = L \mu$ on $K$ is plotted in
Fig.2; the values of $\mu$ obtained by the OTM (Fig.1) have been
used. One can see that our numerical experiments suggest that the
dynamical entropy of the standard map is continuous and monotonous
in $K$, contrary to the discontinuous behavior of LCEs. This is no
wonder, since the dynamical entropy is a more fundamental
quantity.

For orientation, the function $\ln{K \over 2}$ is depicted in the
same Fig.2; this is the well-known logarithmic law derived by
Chirikov~\cite{C78,C79} analytically by means of averaging the
largest eigenvalue of the tangent map in assumption that the
relative measure of the regular component is small.

The downward spikes seen in the $L(K)$ dependence in Fig.2 (and in
Fig.3 also) represent a manifestation of the so-called
``stickiness effect'' immanent to the chaotic Hamiltonian dynamics
in conditions of divided phase space~\cite{S98b}: a chaotic
trajectory may stick for a long time to the borders of the chaotic
domain, where the motion is close to regular, and therefore the
local LCEs are small. Since the computation time is always finite,
the stickiness effect, in the case of deep stickings, leads to
underestimated values of LCEs; see discussion in~\cite{S98b}.

The $L(K)$ and $\mu(K)$ functions are discontinuous and evidently
elude simple analytic representation in broad ranges of $K$. The
case of $h(K)$ is different. With good accuracy, the numerical
dependence at $K$ greater than its critical value can be
approximated by a function which is piecewise linear at moderate
values of $K$ (see Fig.2), and logarithmic with a small power-law
correction at higher values of $K$: $h(K) = \ln {K \over 2} + {1
\over K^2}$, if $K > 4.5$, with accuracy better than $0.01$ in
absolute magnitude. So, the presented numerical data indicates
that the high-$K$ asymptotics of the $h(K)$ function contains a
power-law component, in addition to the well-known logarithmic
one. The same is valid for the $L(K)$ dependence, if one ignores
the small (and local in $K$) distortions of the function due to
accelerator modes and periodic solutions of higher orders.

Let us consider in more detail the accuracy of the presented
results. The LCE values can be effectively verified by controlling
their saturation, taking various values of $n_{it}$. We compare
the maximum LCEs computed taking $n_{it} = 10^7$ with those
computed taking the number of iterations ten times less, $n_{it} =
10^6$. In both cases we average over ten trajectories. One finds
that the difference between the LCEs in these two cases, averaged
over the interval $1 \le K < 2$ (the step in $K$ is $0.01$, so,
$100$ differences are averaged), is equal to only $0.0015$. The
saturation is faster with increasing $K$. Therefore the saturation
at $n_{it} = 10^7$ is practically complete at $K > 1$, and
therefore there are no significant systematic errors in
determination of LCEs at such values of $K$.

The problem of accuracy of computation of $\mu$ is more difficult.
The generic border of chaos in phase space of Hamiltonian systems
is fractal~\cite{UF85,C90}. Umberger and Farmer~\cite{UF85}
conjectured and numerically verified that the coarse-grained
measure of the chaotic constituent of phase space of
two-dimensional area-preserving maps scales with the grid
resolution $\varepsilon$, employed to estimate the measure, as a
power law in the limit $\varepsilon \to 0$:

\vspace{-3mm}

\begin{equation}
\mu_{\varepsilon} = \mu_0 + A \varepsilon^\beta, \label{muf}
\end{equation}

\noindent
where $\mu_0$ is the actual measure, $A$ and $\beta$ are
constants characterizing the border; $A$, $\beta \ge 0$.

Numerical experiments provide values of $\mu_{\varepsilon}$ for a
given $\varepsilon$. So, there are three undetermined quantities
in Eq.~(\ref{muf}): $\mu_0$, $A$, and $\beta$. To obtain their
numerical values one needs to compute $\mu_{\varepsilon}$ at least
thrice, i.~e.\ at three different resolutions $\varepsilon$ of the
grid. Then, the system of three nonlinear equations~(\ref{muf})
can be solved. We take three partitions of phase plane of the
standard map: $2500 \times 2500$, $5000 \times 5000$, and $7500
\times 7500$ pixels; i.~e.\ $\varepsilon = 1/2500$, $1/5000$, and
$1/7500$. As in~\cite{UF85}, the OTM is used; $n_{it} = 10^{10}$.
The minimum $\varepsilon$ and maximum $n_{it}$ used in~\cite{UF85}
were $1/4096$ and $10^8$ respectively. Thus we should expect
better estimates of the chaotic domain measure in the present
study.

In our numerical experiment, we have computed $\mu_0$ and $\beta$
for ten values of $K$ equally spaced in the interval $[1.0, 5.5]$,
and for ten values of $K$ equally spaced in the interval $[1.0,
1.9]$. At all points, the computed value of $\mu_{1/7500}$
($\mu_{\varepsilon}$ at the smallest $\varepsilon = 1/7500$) and
the resulting actual $\mu_0$ value differ by no more than $0.01$;
at $K > 1.5$ they coincide with accuracy of two significant
digits.

Our data for $K=1.1$, $1.2$, and $1.3$ can be compared to results
by Umberger and Farmer~\cite{UF85}, who computed the chaotic
domain measure for these three values of $K$. The difference in
their and our values of $\mu_0$ does not exceed $\approx 0.01$.
Our results on the chaotic domain measure are as well in
reasonable agreement with early estimates by
Chirikov~\cite{C78,C79}. The close proximity of the values of
$\mu_0$ to those of $\mu_{1/7500}$, as well as the agreement of
them with the results~\cite{C78,C79,BGS76}, testify that our
estimates of the chaotic domain measure have the accuracy better
than $0.01$.

The power-law index $\beta$ is related to the fractal dimension
$d_L$ of the set of all chaos borders inside the connected chaotic
domain~\cite{C90}: $d_L = 2 -\beta$. From our data, one has
$\langle \beta \rangle = 0.63 \pm 0.13$, and $d_L \approx 1.37 \pm
0.13$. This agrees well with the theoretical estimate $d_L = 3/2$
by Chirikov~\cite{C90}.

An important constant of the standard map dynamics is the chaotic
domain measure at the critical value of the stochasticity
parameter $K = K_G = 0.971635406\ldots$ (on the critical value,
see e.~g.\ \cite{M92}). Our calculation performed by the same
algorithm as presented above gives $\mu(K_G) \approx 0.463$. The
contribution of the chaotic domain around the integer resonance to
this quantity is $3.5$ times greater than that of the half-integer
one ($0.463 \approx 0.359 + 0.103$). The calculated values of the
parameter $\beta$ in these domains are equal to $\approx 0.53$ and
$\approx 0.49$ respectively; so, the border fractal dimension $d_L
= 2 -\beta$ in the critical case $K = K_G$ is particularly close,
as one could expect, to the theoretical estimate $d_L = 3/2$ by
Chirikov~\cite{C90}.

\section{Estimation of Chirikov's constant}
\label{echc}

The value of Chirikov's constant can be found by calculating the
average of the local maximum LCE over the chaotic layer of
map~(\ref{sm1}) in the limit $\lambda \to \infty$. The local LCE
must be taken with weight directly proportional to the time that
the trajectory spends in a given part of the layer; this time is
directly proportional to the local relative measure of the chaotic
component. Therefore one has the following formula

\begin{equation}
C_h = \lim_{\lambda \to \infty} \frac{\int\limits_0^{y_b} \tilde
L_{sx}(y) \tilde \mu_{sx}(y) \, dy}{\int\limits_0^{y_b} \tilde
\mu_{sx}(y) \, dy}, \label{chint}
\end{equation}

\noindent where $y_b = \lambda / K_G$ is the value of $y$ at the
border of the layer, $\tilde L_{sx}(y)$ is the local (with respect
to $y$) value of the maximum LCE of the separatrix map, and
$\tilde \mu_{sx}(y)$ is the local chaos measure. The tilde cap
marks that the quantities are local. This formula is valid in the
limit $\lambda \to \infty$, since only in this limit one can
reduce the sum over all integer resonances inside the layer to an
integral. What is more, the formula $y_b = \lambda / K_G$
(see~\cite{C90,S98a}) is accurate also only in this limit.

By means of the substitution $y=\lambda/K$ we introduce a new
independent variable $K$, which is nothing but the value of the
stochasticity parameter of the standard map locally approximating
the separatrix map. The accuracy of the approximation improves
with increasing $\lambda$~\cite{C78,C79}, i.~e.\ the local
characteristics of the chaotic layer converge to those of the
standard map locally approximating the motion: $\tilde
L_{sx}(y=\lambda/K) \to L(K)$ and $\tilde \mu_{sx}(y=\lambda/K)
\to \mu(K)$, where $L(K)$ and $\mu(K)$ are the maximum LCE and the
measure of the main connected chaotic domain of the standard map
in function of $K$. The dependence on $\lambda$ in the limit
$\lambda \to \infty$ is eliminated, and Eq.~(\ref{chint}) is
reduced to the final form:

\begin{equation}
C_h = \frac{K_G}{\sigma} \int\limits_{K_G}^\infty L(K) \mu(K) \,
\frac{dK}{K^2} \, , \label{chint1}
\end{equation}

\noindent where the quantity

\begin{equation}
\sigma = \lim_{\lambda \to \infty} y_b^{-1} \int\limits_0^{y_b}
\tilde \mu_{sx}(y) \, dy = K_G \int\limits_{K_G}^\infty \mu(K) \,
\frac{dK}{K^2} \label{sg}
\end{equation}

\noindent has the meaning of ``porosity'' of the chaotic layer.
This is the ratio of the area of the chaotic component to the
total area of the layer bounded by its external borders; the
quantity $1-\sigma$ is nothing but the total relative area of all
regular islands inside the layer.

For $K \in [K_G, 10]$, we integrate Eqs.~(\ref{chint1}, \ref{sg})
numerically; the functions $L(K)$ and $\mu(K)$ are taken in
numerical form, as presented in Figs.~1 and 2. The remainders for
$K > 10$ are calculated analytically, with $h(K) = L(K) \mu(K)$
set equal to $\ln {K \over 2} + {1 \over K^2}$ (as established
above), and $\mu(K)$ set equal to unity. Adopting accuracy of two
significant digits, one has $C_h = 0.80$, $\sigma = 0.78$.

Our value of $C_h$ differs significantly from the value of $0.663$
got by Chirikov by integration of the dynamical entropy of the
standard map in~\cite{C78,C79} (where $C_h$ is designated as
$h_W$). This deviation is due to the sparsity of the numerical
data obtained more than twenty years ago, as well as to ignoring
the porosity of the chaotic layer in an approximate calculation
in~\cite{C78,C79}.

As discussed in the previous Section, the numerical dependence
$\mu(K)$ is less certain than $L(K)$. Hence its uncertainty is the
most likely source of errors in estimating $C_h$. The estimated
error in determination of $\mu(K)$ does not exceed $0.01$ (see
above). At high values of $K$ (at $K > 6$), the deviations are
much less than $0.01$, because $\mu(K)$ rapidly converges to
unity. To estimate the accuracy of the obtained value of $C_h$, we
recompute this value substituting $\mu(K) \pm 0.01$ instead of the
original $\mu(K)$ in Eqs.~(\ref{chint1}) and~(\ref{sg}) at $K \le
10$ (if $\mu(K) + 0.01 > 1$ we set $\mu(K) = 1$, of course). We
find that both these negative and positive shifts in $\mu(K)$
change the resulting value of $C_h$ by no more than $0.004$.
Therefore, if one takes three significant digits in $C_h$, the
result is $C_h = 0.801 \pm 0.004$ in the described sense. The
deviations in $\sigma$ are greater: $\sigma = 0.780 \pm 0.009$.
Finally, rounding up, we conclude that the accuracy of our
estimate $C_h = 0.80$ is presumably better than $0.01$.

One can verify this estimate of Chirikov's constant by means of a
straightforward computation of the maximum LCE of the separatrix
map at high values of $\lambda$. The $\lambda$ dependence of the
maximum LCE of the separatrix map is shown in Fig.3. It has been
obtained by a numerical experiment with map~(\ref{sm1}). The use
of one and the same designation $L$ for the LCE in the both cases
of the standard and separatrix maps should not cause confusion.
The resolution (step) in $\lambda \in [0, 10]$ is equal to $0.05$.
At each step in $\lambda$, the values of $L$ have been computed
for 100 values of $c$ equally spaced in the interval $[0, 2 \pi]$.
The number of iterations for each trajectory is $n_{it} = 10^7$
for $\lambda \in [0, 1)$ and $n_{it} = 10^8$ for $\lambda \in [1,
10]$ (the saturation time for numerical estimates of LCEs
increases with $\lambda$, and therefore more computational time is
needed to get reliable values of LCEs at high values of
$\lambda$). This is sufficient to saturate the computed values of
$L$; see below. At each step in $\lambda$ we find the value of $c$
corresponding to the minimum width of the layer (the case of the
least perturbed border), and plot the value of $L$ corresponding
to this case. The case of the least perturbed border is generic in
applications, in the sense that strong perturbations of the border
are local in $c$. Note that the parameter $c$ is related to the
amplitude of the periodic perturbation in Hamiltonian~(\ref{h}) at
a given value of $\lambda$ by Eq.~(\ref{c}).

Fig.3 also presents approximation of the observed dependence by
the rational function

\begin{equation}
L(\lambda) = {b + c \lambda \over 1 + a \lambda} \label{lsxap}
\end{equation}

\noindent with $b$ set to zero in order that $L(0) = 0$. The
resulting values of the parameters and their standard errors are:
$a = 2.097 \pm 0.033$, $b = 0$, and $c = 1.691 \pm 0.024$.

Chirikov's constant is given by the limit $L(\lambda \to \infty)$;
so, according to the described numerical experiment, $C_h \approx
0.806$, in good agreement with the result presented above ($C_h
\approx 0.801$).

The integration time we used is sufficient for effective
saturation of the computed LCE in the given interval of $\lambda$.
Indeed, setting $n_{it} = 10^7$ for the whole interval $\lambda
\in [0, 10]$ gives the resulting $C_h \approx 0.808$, i.~e.\ the
resulting $C_h$ value is negligibly different from the value
obtained with $n_{it}$ raised to $10^8$ at $\lambda \in [1, 10]$.

Variation of the parameter $c$ in map~(\ref{sm1}) produces a
scatter in the computed values of LCE, due to emergence and
disappearance of marginal resonances at the border of the chaotic
layer (on the marginal resonances see~\cite{S98a}). Let us prove
that the LCE scatter tends to zero in the limit $\lambda \to
\infty$; in other words, the limit of LCE is one and the same for
all (though sufficiently small, for the map description to be
valid) amplitudes of perturbation.

The largest variations are conditioned by integer marginal
resonances. The $y$ coordinates of the centers of the integer
resonances satisfy the relation $y_{i+1}/y_i = \exp(2 \pi/
\lambda)$, where $i$ is the number of the resonance. This relation
follows from the second line of map~(\ref{sm1}). At the border of
the layer $y \approx \lambda$ (see~\cite{C78,C79}); therefore in
the case of $\lambda \gg 1$ the distance between the centers of
two consecutive integer resonances at the border is $\Delta y
\approx 2 \pi$. The relative local measure $\mu_{marg}$ of the
chaotic component associated with the separatrices of a marginal
resonance depends on the value of the parameter $K$ of the
standard map locally approximating the motion near the marginal
resonance; the maximum value is $\approx 0.46$ at $K = K_G$ (see
the concluding paragraph of Section~\ref{lede}), because any
marginal resonance has $K < K_G$. This relative measure referred
to the total chaotic measure of the layer is less than
approximately $(\mu_{marg} \cdot 2 \pi) / (\sigma \cdot \lambda)
\approx 3.7/\lambda$. The largest value of the local LCE at the
border is again associated with $K = K_G$; it equals approximately
$0.11$ (see Fig.2). So, the contribution of the chaotic layer of
the marginal resonance to the total value of the maximum LCE over
the whole layer varies from zero up to $\approx 0.4/\lambda$,
depending on the prominence of the marginal resonance, i.~e.\ on
the local value of $K$ at the border. This contribution tends to
zero with $\lambda \to \infty$, and therefore the value of
Chirikov's constant does not depend on the second parameter of the
separatrix map, $c$, and consequently on the amplitude of the
periodic perturbation.

\section{Conclusions}

Exploiting our high-precision data on the functions $h(K)$ and
$\mu(K)$, we have calculated the value of Chirikov's constant
$C_h$~--- the least upper bound for the maximum Lyapunov exponent
of the separatrix map. This quantity is nothing but the maximum
Lyapunov exponent (referred to the mean half-period of phase
libration, or, equivalently, to the mean period of phase rotation)
of the motion in the chaotic layer of a nonlinear resonance
subject to symmetric periodic perturbation, in the limit of
infinitely high frequency of the perturbation. We have shown that
the value $C_h$ does not depend on the second parameter of the
separatrix map (or, equivalently, on the amplitude of the
perturbation).

The newly derived value of $C_h$ is $0.80$, with precision
presumably better than $0.01$. The knowledge of this constant is
important for correct analytical estimation of the value of the
maximum Lyapunov exponent of the chaotic motion of a Hamiltonian
system allowing description in the perturbed pendulum model.

The author is thankful to B.\,V.~Chirikov and V.\,V.~Vecheslavov
for valuable discussions. This work was supported by the Russian
Foundation for Basic Research (project number 03-02-17356).

\begin{figure}[ht!]
\begin{center}
(a)~\includegraphics[width=10cm]{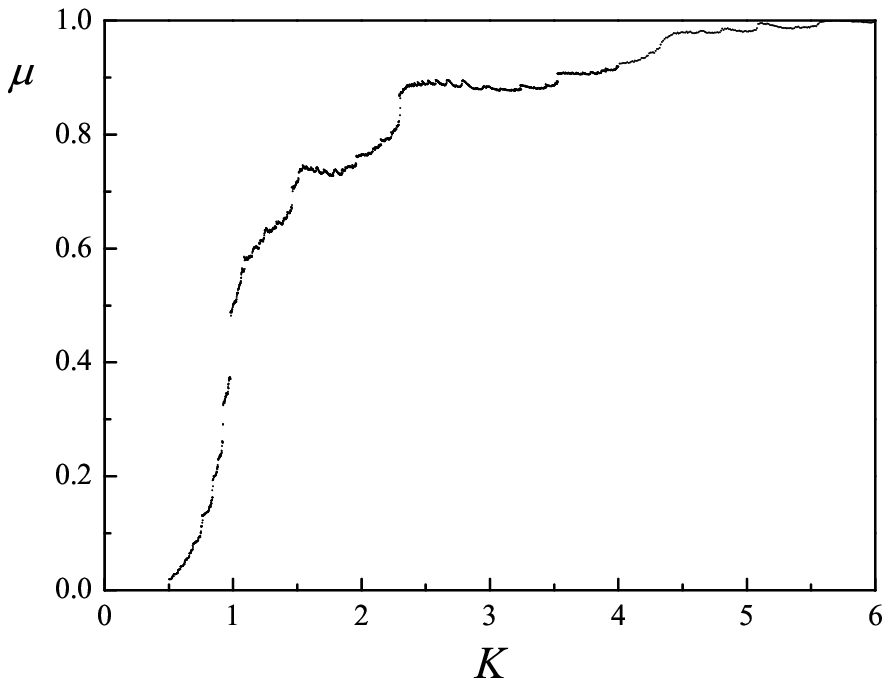} \\
(b)~\includegraphics[width=10cm]{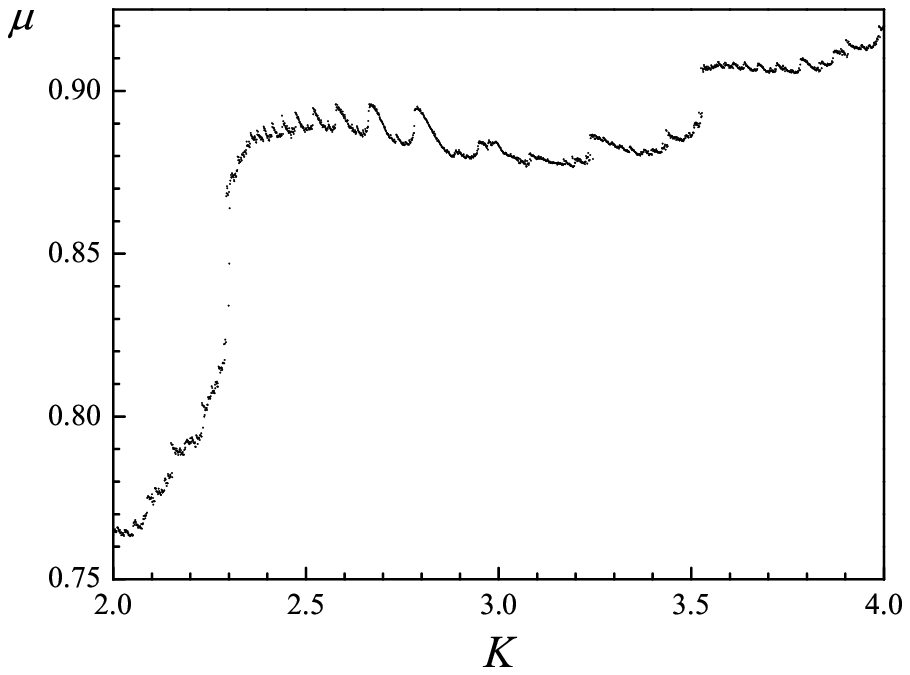}
\end{center}
\caption{\small The dependence $\mu(K)$~(a); a detail
enlarged~(b)} \label{fig1}
\end{figure}

\begin{figure}[ht!]
\begin{center}
\includegraphics[width=12cm]{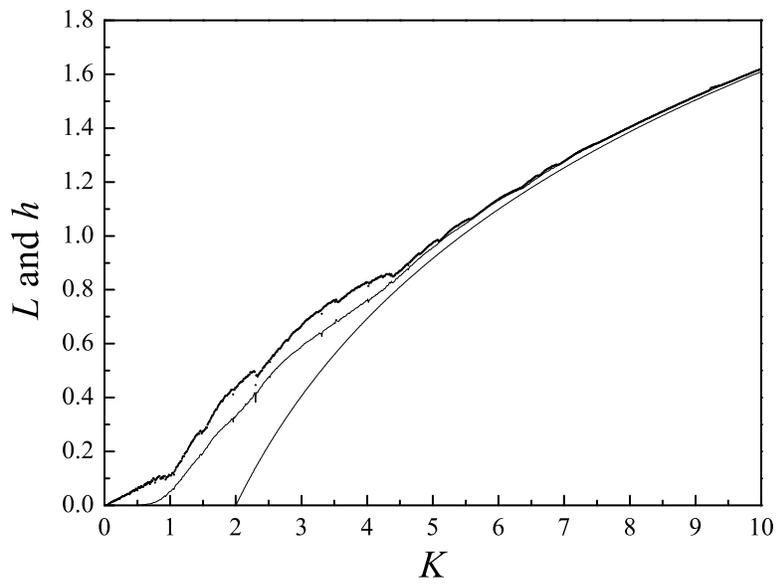}
\end{center}
\caption{\small $L(K)$ (the uppermost curve), $h(K)$, and
$\ln{\frac{K}{2}}$ in a broad range of $K$} \label{fig2}
\end{figure}

\begin{figure}[ht!]
\begin{center}
\includegraphics[width=12cm]{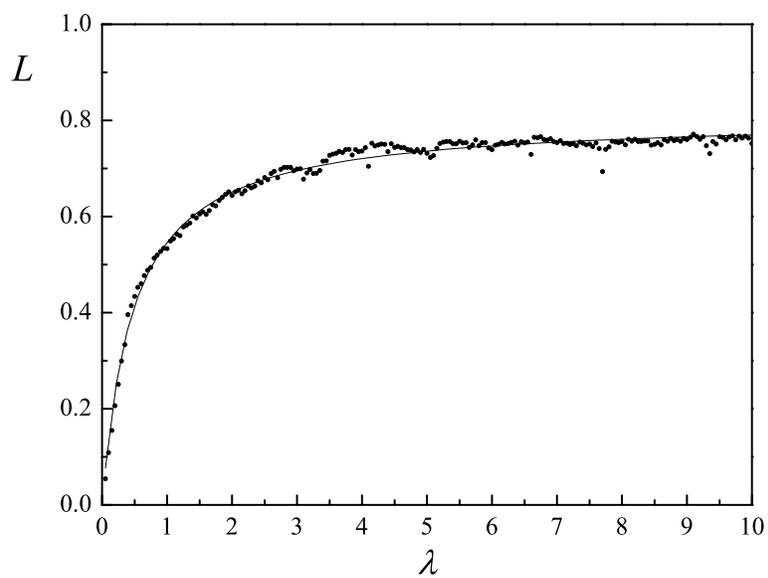}
\end{center}
\caption{\small The dependence $L(\lambda)$ for the separatrix map
(the case of the least perturbed border of the chaotic layer) and
its rational approximation} \label{fig3}
\end{figure}

\end{document}